\documentclass[12pt,journal,final,onecolumn]{IEEEtran}

\usepackage[dvips]{graphicx}
\usepackage{epsfig}
\usepackage[cmex10]{amsmath}
\usepackage{amssymb}
\usepackage{amsthm}
\usepackage{amsfonts}
\usepackage{bm}
\usepackage{cite}
\usepackage[svgnames]{xcolor} % load before pstricks
\usepackage{pstricks,pst-node,pst-plot,pstricks-add}
\usepackage[tight,footnotesize]{subfigure}
\usepackage{algorithm}
\usepackage{algpseudocode}

\graphicspath{{figs/}}

\newcommand{\mc}[1]{\mathcal{#1}}

\newcommand{\defeq}{\mathrel{\triangleq}}

\newcommand{\N}{\mathbb{N}}

\newcommand{\abs}[1]{\lvert{#1}\rvert}
\newcommand{\card}[1]{\abs{#1}}

\newcommand{\iid}{i.\@i.\@d.\ }

\newtheorem{innercustomtheorem}{Theorem}
{\renewcommand\theinnercustomtheorem{#1}\innercustomtheorem}%
{\endinnercustomtheorem}

\theoremstyle{definition}

\newtheorem{egdummy}{Example}
\newenvironment{example}[1][]%
{%
    \begin{egdummy}[#1]%
    \upshape%
}%
{%
    \qed%
    \end{egdummy}%
}

\newtheoremstyle{myremark}%
{\topsep}{\topsep}{\normalfont}{\parindent}{\itshape}{:}{ }{}

\theoremstyle{myremark}

\newcommand\shortintertext[1]{%
\ifvmode\else\\\@empty\fi
\noalign{%
\penalty0%
\vbox{\mathstrut}%
\penalty10000%
\vskip-\baselineskip
\penalty10000%
\vbox to 0pt{%
\normalbaselines
\ifdim\linewidth=\columnwidth
\else
\parshape\@ne
\@totalleftmargin\linewidth
\fi
\vss
\noindent#1\par}%
\penalty10000%
\vskip-\baselineskip}%
\penalty10000}

\begin{document}

\title{Coded Caching for Delay-Sensitive Content}

\author{Urs Niesen and Mohammad Ali Maddah-Ali%
    \thanks{The authors are with Bell Labs, Alcatel-Lucent. Emails:
        urs.niesen@alcatel-lucent.com,
        mohammadali.maddah-ali@alcatel-lucent.com.}%
}

\maketitle

\begin{abstract}
    Coded caching is a recently proposed technique that achieves
    significant performance gains for cache networks compared to uncoded
    caching schemes. However, this substantial coding gain is attained
    at the cost of large delivery delay, which is not tolerable in
    delay-sensitive applications such as video streaming. In this paper,
    we identify and investigate the tradeoff between the performance
    gain of coded caching and the delivery delay. We propose a
    computationally efficient caching algorithm that provides the gains
    of coding and respects delay constraints. The proposed algorithm
    achieves the optimum performance for large delay, but still offers
    major gains for small delay. These gains are demonstrated in a
    practical setting with a video-streaming prototype.
\end{abstract}

\section{Introduction}
\label{sec:intro}

Video on demand drives network traffic growth today, and this trend is
expected to continue until the end of the decade~\cite{cisco14}. Indeed,
some studies predict that video traffic will grow by a factor 12 between
2010 and 2020~\cite{lucent12}.  Video traffic is time varying in nature,
with a significant peak during the evenings. This characteristic burdens
service providers with the cost of having to expand their networks to
handle increasing peak traffic even though the network remains
underutilized during off-peak times.

Caching is a key technique to alleviate this problem by smoothing
traffic in time and across the network. Caching uses memories
distributed throughout the network to duplicate popular content during
off-peak times. During peak traffic times, these memories are then used
to deliver requested content with less stress on the network.

The core idea behind conventional caching schemes
\cite{dowdy82,sleator85,almeroth96,dan96,korupolu99,meyerson01,baev08,borst10}
is to deliver part of the content locally from nearby caches.  Recently
\cite{maddah-ali12a,maddah-ali13} proposed instead to use \emph{coded
caching}. Unlike conventional uncoded schemes, coded caching creates
linear combinations of chunks of content requested by different users.
The combinations are formed such that each user can recover his
requested chunk using the content of a nearby cache.  Therefore the
caches are used not only to deliver parts of the content locally, but
also to create coded multicasting opportunities.
\cite{maddah-ali12a,maddah-ali13} shows that, for some basic network
structures, coded caching can yield significantly better performance
than uncoded caching. This result demonstrates that widely-used caching
algorithms such as least-frequently used (LFU) or least-recently used
(LRU) can be significantly suboptimal for cache
networks~\cite{niesen13,pedarsani13}.

The substantial gain of coded caching is achieved at the cost of
increased delay during content delivery. However, many applications,
such as video on demand, are delay sensitive, and consequently this
added delay resulting from  coding is problematic. This raises the
question of how much of the coded caching gain can be obtained within
some fixed delay constraint.

In this paper, we initiate the investigation of this tradeoff between
coded caching gain and delivery delay. We develop a computationally
efficient caching scheme that effectively exploits coding opportunities
while respecting delivery-delay constraints. The proposed scheme
achieves the optimum performance suggested
by~\cite{maddah-ali12a,maddah-ali13} in the large-delay limit, while
still offering a significant gain for delay-sensitive applications.  A
video-streaming prototype implementing these ideas confirms the
practicality and gains of the proposed scheme.

\section{Problem Formulation}
\label{sec:problem}

We consider an origin server connected through a network to $K$ edge
caches. The server stores a collection of videos, split into a number of
symbols. Each symbol corresponds to a chunk of data of constant
size, say between $1$ and $10$ kB, plus a header providing a unique
identifier for the symbol. To each edge cache are attached a number of
users, and for simplicity we assume that each user can access only a
single edge cache.

In a \emph{placement phase}, occurring during a time of low network
load, each cache independently prefetches every symbol independently at
random with some probability $p$, with $p\in[0,1]$ chosen such that
memory constraints at the users are satisfied as proposed
in~\cite{maddah-ali13}. Therefore, approximately a fraction $p$ of each
video is stored at each cache. We assume that the server knows which
content symbols are stored in which cache (see the discussion in
Section~\ref{sec:implementation} for how this can be achieved in
practice).  In a later \emph{delivery phase}, occurring during a time of
high network load, the users attached to the caches issue a sequence of
requests, each request for one content symbol. 

The server can represent each such request by a triple $(k;\, \mc{S};\,
t)$, where $k\in\{1, 2, \dots, K\}$ is the cache handling the user
requesting the symbol, $\mc{S}\subset\{1, 2, \dots, K\}$ is the subset
of caches that have stored the requested symbol, and $t$ is the deadline
by which this request needs to be served. For ease of notation, we will
assume that $t\in\N$ is an integer, measured in some arbitrary unit, say
milliseconds, and that time starts at $t=0$.  For the same reason, we
sometimes write $(k;\, \mc{S})$ for $(k;\, \mc{S};\, t)$ when the value
of $t$ is immaterial.

Note that this notation suppresses the identifier of the actual symbol
requested by the user, as this identifier is not needed for most of the
subsequent analysis. Observe further that each request $(k; \mc{S})$
satisfies the property $k\notin\mc{S}$, because otherwise this request
could be served from the user's local cache without contacting the server.

\begin{example}
    \label{eg:requests}
    Consider a user attached to cache one requesting a video in a
    streaming application. At time $t=0$, the user issues a sequence of
    requests for the first few consecutive symbols of the video. The
    first request might have a tight deadline, say of $t=1$; the second
    request might have a less stringent deadline of $t=5$. Assume that the
    first requested symbol is available at caches two and three and that
    the second requested symbol is available only at cache two. The two
    requests are then described by\footnote{$(1;\, 2, 3;\, 1)$ is used
        as notational shorthand for $(1;\, \{2, 3\};\, 1)$.  Similar
    simplified notation is used throughout.} $(1;\, 2, 3;\, 1),\ (1;\,
    2;\, 5)$. 
\end{example}

The server responds to these requests by sending multicast packets to
all the $K$ edge caches associated with this server, and our aim is to
minimize the bandwidth required by the server. The key observation is
that the cached symbols can be used to create coded multicasting
opportunities, in which a single coded packet is simultaneously useful
for several users with different demands~\cite{maddah-ali12a}. 

\begin{example}
    \label{eg:coding}
    Consider the sequence of requests
    \begin{equation*}
        (1;\, 2, 3),\ (2;\, 1),\ (2;\, 1, 3),\ (3;\, 1, 2),\ (1;\, 2)
    \end{equation*}
    issued by three distinct users. The first request is via cache one
    requesting a symbol available at caches two and three, the second
    request is via cache two for a symbol available at cache one, and so
    on.

    For concreteness, assume that the five requested symbols are called
    $A$, $B$, $C$, $D$, $E$. The server could satisfy all five requests
    by transmitting $A,\ B,\ C,\ D,\ E$.  However, by making use of the
    cache contents, the server can do better. Note that symbol $B$
    requested via cache two is available at cache one. Similarly, symbol
    $E$ requested via cache one is available at cache two. Hence, from
    the single multicast transmission $B\oplus E$  and using their
    stored symbols, both caches can recover their requested symbol. Here
    $\oplus$ denotes the bitwise XOR operation.  By an analogous
    argument, caches one, two, and three can recover their respective
    requested symbols $A$, $C$, and $D$ from the single multicast
    transmission $A\oplus C\oplus D$. Thus, the server can satisfy all
    five requests by just two coded multicast transmissions, $B\oplus E$
    and $A\oplus C\oplus D$. Because the XOR of several symbols has the
    same size as a single symbol, this approach reduces the required
    bandwidth out of the server by a factor $5/2$.
\end{example}

From Example~\ref{eg:coding}, we see that two requests $(k_1;\, \mc{S}_1;\,
t_1)$ and $(k_2;\, \mc{S}_2;\, t_2)$ can be \emph{merged}, meaning that they
can both be satisfied by a single coded multicast transmission from the
server, if $k_1\subset\mc{S}_2 $ and $k_2\subset\mc{S}_1$. We denote the
merged request by $\bigl( \{k_1, k_2\};\, \mc{S}_1\cap \mc{S}_2;\,
\min\{t_1, t_2\}\bigr)$. This notations indicates that this new request
is to be sent to both caches $k_1$, $k_2$, has a deadline of $\min\{t_1,
t_2\}$, and the symbols stored in both caches are $\mc{S}_1\cap \mc{S}_2$.

Consider next two such merged requests $(\mc{K}_1;\, \mc{S}_1;\, t_1)$
and $(\mc{K}_2;\, \mc{S}_2;\, t_2)$. These merged requests can be
further merged, if
\begin{equation}
    \label{eq:mergeable}
    1.\ \mc{K}_1 \subset\mc{S}_2 \qquad\text{and}\qquad 
    2.\ \mc{K}_2 \subset\mc{S}_1.
\end{equation}
The corresponding merged request is
\begin{equation*}
    \bigl(\mc{K}_1\cup \mc{K}_2;\, \mc{S}_1\cap \mc{S}_2;\, \min\{t_1, t_2\}\bigr).
\end{equation*}

Whenever the current time is equal to the time $t$ of the merged
request, the server needs to respond to this request by transmitting the
corresponding coded multicast packet.  In order to minimize the required
bandwidth out of the server, our goal is to merge requests as much as
possible, thereby minimizing the number of transmitted coded multicast
packets. The reduction in server bandwidth resulting from coded
multicasting compared to uncoded transmission is the \emph{global coding
gain}, defined as the ratio between the bandwidth of the uncoded scheme
and the coded scheme. We adopt this global coding gain as our figure of
merit in this paper. 

The goal is thus to design rules to merge user requests such as to
maximize this global coding gain. Because symbols are requested by the
users sequentially and need to be served within a deadline, this merging
needs to be performed online, i.e., among the requests currently pending
at the server.  Furthermore, because the queue of pending requests can be
quite large as illustrated by the next example, the merging algorithm
needs to be computationally efficient.

\begin{example}
    \label{eg:length}
    For an HTTP adaptive streaming application with $100$ users,
    video rate of $400$ kb/s, delay tolerance of $2$ s, and symbol size of
    $10$ kB, the server queue will have approximately $10^3$ unmerged
    requests at any given time.
    For a progressive download video on-demand application with $600$ users,
    video rate of $1800$ kb/s, delay tolerance of $30$ s, and symbol size of
    $4$ kB, the server queue will have approximately $10^6$ unmerged
    requests at any given time.
    For a complete download video on-demand application with $1000$ users,
    video rate of $2400$ kb/s, delay tolerance of $3600$ s, and symbol size of
    $1$ kB, the server queue will have approximately $10^9$ unmerged
    requests at any given time.
\end{example}

\section{Request-Merging Rules}
\label{sec:main}

We now present several request-merging rules and compare their
performance. We start with two examples to motivate the new proposed
rule introduced below.

\emph{First-Fit Rule}: Consider an ordered sequence of (potentially
merged) requests $(\mc{K}_1;\, \mc{S}_1;\, t_1)$, $(\mc{K}_2;\,
\mc{S}_2;\, t_2)$, $\dots$, $(\mc{K}_{L-1};\, \mc{S}_{L-1};\, t_{L-1})$
with $t_1 \leq t_2 \leq \dots \leq t_{L-1}$. Assume that a new request
for a single symbol $(k_L;\, \mc{S}_L;\, t_L)$ arrives at the server. In
the first-fit rule, the server traverses its queue starting from
$\ell=1$ to find the first (tightest deadline) request $(\mc{K}_\ell;\,
\mc{S}_\ell;\, t_\ell)$ that can be merged with the new request, i.e.,
such that the two conditions in~\eqref{eq:mergeable} hold.  The newly
merged request is
\begin{equation*}
    \bigl(\mc{K}_\ell\cup \{k_L\};\, \mc{S}_\ell\cap \mc{S}_L;\, \min\{t_\ell, t_L\}\bigr).
\end{equation*}
If no such request can be found in the queue, the new request $(k_L;\,
\mc{S}_L;\, t_L)$ is appended to the queue. This type of
first-fit rule has a long history; most relevant to this work is that it
was used in \cite{grimmett75} in the context of coloring random graphs.

The first-fit rule falls into the class of \emph{sequential} merging
rules. These rules keep a list of merged requests at the server.
Whenever a new request arrives, these algorithms traverse this list
sequentially starting from the request with the tightest deadline and
try to merge the new request with one already in the queue. The decision
to merge with a queued request $(\mc{K}_\ell;\, \mc{S}_\ell;\, t_\ell)$
is made based only on the information in that request. 

Sequential merging rules have several advantages. First, because each
queued request is considered only once, and because each such
consideration is based on only information in the request and hence
takes $O(1)$ time to evaluate, the computational complexity to insert
one new request into the server queue is $O(L)$, where $L$ is the queue
length. Second, for the same reasons, such rules can be easily
parallelized. Indeed, several threads can sequentially traverse the
queue in parallel, each handling the merging operation of one new
request. Thus, sequential merging rules are computationally efficient.

We next introduce a new sequential merging rule.

\emph{Perfect-Fit Rule}: This is a sequential merging rule, in which a
new request $(k_L;\, \mc{S}_L;\, t_L)$ is merged with the first queued
request $(\mc{K}_\ell;\, \mc{S}_\ell;\, t_\ell)$ satisfying the
following three conditions:
\begin{equation*}
    1.\ k_L \in\mc{S}_\ell \qquad
    2.\ \mc{K}_\ell \subset\mc{S}_L \qquad
    3.\ \mc{S}_\ell\cup\mc{K}_\ell = \mc{S}_L\cup k_L
\end{equation*}
The first two conditions ensure that the two requests under
consideration can in fact be merged. The third condition stipulates that
the merging should be optimal in terms of the size of the intersection
of the cache availability $\mc{S}_\ell\cap\mc{S}_L$ of the merged
request. The relevance of this third condition will be discussed in
more detail below.

The first-fit and perfect-fit rules can be seen to be approximately
optimal in terms of minimizing server bandwidth in certain regimes. For
the large-queue regime, $L\to\infty$, it is shown in~\cite{maddah-ali13}
that a nonsequential version of the perfect-fit rule is optimal. Indeed,
as $L\to\infty$, with high probability there will be a queued request
that satisfies all three merging conditions, and as we will see below,
the third condition ensures that the server bandwidth is minimized. For
the small-queue regime, $L\to 0$, and assuming a large enough number of
caches $K$, results on online coloring of random graphs~\cite{grimmett75}
suggest that the first-fit rule is close to optimal.

The regime of most interest is clearly when the queue length is in
between those two extremes (see Example~\ref{eg:length} in
Section~\ref{sec:problem}). We next introduce a novel family of
sequential merging rules for this intermediate regime. Before we
describe these rules in detail, we need some additional notation.

Let $(\mc{K}_1;\, \mc{S}_1;\, t_1)$ and $(\mc{K}_2;\, \mc{S}_2;\, t_2)$
be two mergeable requests, and consider their merged version.  Merging
these two requests has two effects on the system, one beneficial and one
detrimental. The beneficial effect is that both requests can now be
served with a single coded multicast transmission. The detrimental
effect is that it is now harder for other requests to further merge with
this merged version. The following example highlights the fundamental
tension between these two effects.

\begin{example}
    \label{eg:misfit1}
    Consider the sequence of requests $(1;\, 2, 3)$,\ $(2;\, 1)$,\
    $(2;\, 1, 3)$. The request $(1;\, 2, 3)$ could be merged with either
    $(2;\, 1)$, creating the merged request $(1, 2;\, )$, or with $(2;\,
    1, 3)$, creating the merged request $(1, 2;\, 3)$. Both of these
    merged requests correspond to two users, however, the first one
    cannot accept any further requests whereas the second one can accept
    an additional request such as $(3;\, 1, 2)$. 
    
    Assume while traversing the queue to insert the new request $(1;\,
    2, 3)$ we encounter the already queued request $(2;\, 1)$. We can
    either merge the two requests at the cost of destroying merging
    opportunities for future requests, or we can decide not to merge
    them at the cost of potentially being unable to merge the current
    request at all. Resolving this fundamental tension is at the core
    of a good request-merging rule.
\end{example}

\begin{figure}[htbp]
    \centering 
    \includegraphics{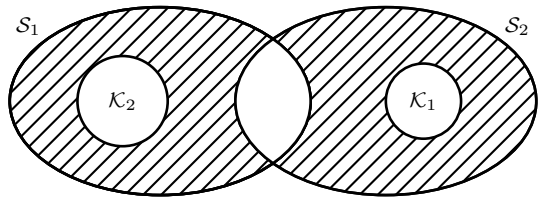} 

    \caption{ The cardinality of the shaded area is the misfit between
        the two mergeable requests $(\mc{K}_1;\, \mc{S}_1)$ and
        $(\mc{K}_2;\, \mc{S}_2)$.}
    \label{fig:misfit}
\end{figure}

To capture this tension, we introduce the following \emph{misfit function}
$\rho(\cdot, \cdot)$. For two requests $(\mc{K}_1;\, \mc{S}_1)$ and
$(\mc{K}_2;\, \mc{S}_2)$, the misfit is given by
\begin{equation*}
    \rho\bigl((\mc{K}_1;\, \mc{S}_1),\, (\mc{K}_2;\, \mc{S}_2) \bigr) 
    \defeq \card{\mc{S}_1\setminus (\mc{S}_2\cup\mc{K}_2)} 
    +\card{\mc{S}_2\setminus (\mc{S}_1\cup\mc{K}_1)}
\end{equation*}
if $\mc{K}_1\subset\mc{S}_2$ and $\mc{K}_2\subset\mc{S}_1$ (ensuring
that the requests are mergeable), and the misfit is equal to $\infty$
otherwise, as is depicted in Fig.~\ref{fig:misfit}. The misfit between
two requests takes values in the set $\{0, 1, \dots, K-3, K-2,
\infty\}$, where $K$ is the number of caches. 

This definition is motivated as follows. Assume we merge $(\mc{K}_1;\,
\mc{S}_1)$ and $(\mc{K}_2;\, \mc{S}_2)$. The cached symbols $\mc{K}_1$
in $\mc{S}_2$ and $\mc{K}_2$ in $\mc{S}_1$ are thus useful to enable the
merging. Furthermore, the symbols in $\mc{S}_1\cap\mc{S}_2$ will be
useful to further merge the already merged requests. On the other hand,
the symbols in $(\mc{S}_1\setminus
(\mc{S}_2\cup\mc{K}_2))\cup(\mc{S}_2\setminus (\mc{S}_1\cup\mc{K}_1))$
are not used for either purpose and will be wasted by merging the two
requests.  The misfit function $\rho(\cdot,\cdot)$ thus measures the
number of these wasted symbols.

\begin{example}
    \label{eg:misfit2}
    Consider the same sequence of requests as in
    Example~\ref{eg:misfit1}. Here we have the misfit values
    \begin{align*}
        \rho\bigl( (1;\, 2, 3),\, (2;\, 1) \bigr) & = 1 + 0 = 1, \\
        \rho\bigl( (1;\, 2, 3),\, (2;\, 1, 3) \bigr) & = 0 + 0 = 0,
    \end{align*}
    consistent with our intuition that the second merged
    request is a better fit than the first merged request. 
\end{example}

The perfect-fit rule can now be equivalently described as the sequential
rule that merges the new request with the first queued request with zero
misfit.  This scheme performs well for long queues, when we are likely
to find queued requests that can be merged with the new request without
wasting any symbols. Similarly, the first-fit rule can be equivalently
described as the sequential rule that merges the new request with the
first queued request with finite misfit. This rule performs well for
short queues, where any chance to merge requests cannot be missed,
irrespective of the number of wasted symbols. 

This suggests the introduction of a general class of merging rules,
which we term \emph{$\tau$-fit threshold rule}. 

\emph{$\tau$-Fit Threshold Rule}: This is a
sequential merging rule, in which a new
request $(k_L;\, \mc{S}_L;\, t_L)$ is merged with the first queued request
$(\mc{K}_\ell;\, \mc{S}_\ell;\, t_\ell)$ such that 
\begin{equation*}
    \rho\bigl( (k_L;\, \mc{S}_L),\,
    (\mc{K}_\ell;\, \mc{S}_\ell) \bigr) \leq \tau.
\end{equation*}

\begin{figure}[htbp]
    \centering 
    \hspace{-1.9cm}\includegraphics{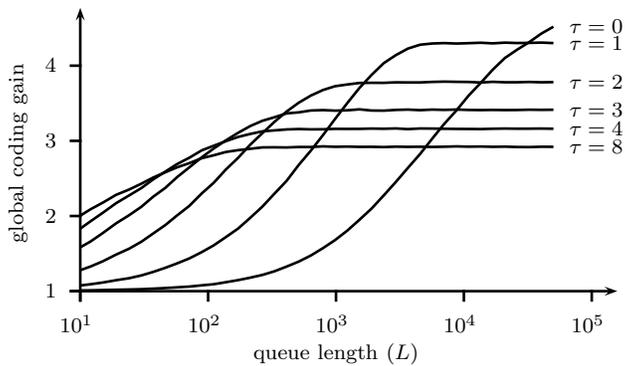}
    \caption{Scaling of the global coding gain (i.e., the bandwidth
        ratio of uncoded to coded transmission) as a function
        of queue length $L$ for $\tau$-fit threshold merging
        rule for different values of $\tau$. In the figure, the number
    of caches is fixed to $K=10$ and the cache probability to $p=0.5$.}
    \label{fig:scalingL}
\end{figure}

We thus see that the first-fit rule is a $K-2$-fit threshold rule and
that the perfect-fit rule is a $0$-fit threshold rule. As $\tau$ varies
from $K-2$ to $0$, the $\tau$-fit threshold rule describes a family of
sequential merging rules. The optimal value $\tau^\star$ of the
threshold $\tau$ depends strongly on the parameters $K$ and $L$, and to
a lesser extent on the parameter $p$. We expect the value of the optimal
threshold $\tau^\star$ to be decreasing with the queue length $L$; when
$L\ll K$, $\tau^\star = K-2$, when $L \gg K 2^K$, $\tau^\star = 0$.
This is indeed the case, as is illustrated in Fig.~\ref{fig:scalingL}.

More formally, in the limit as the queue length $L\to\infty$, a
nonsequential algorithm is shown in~\cite{maddah-ali13} to achieve a
global coding gain of
\begin{equation}
    \label{eq:limit}
    \frac{Kp}{1-(1-p)^K}.
\end{equation}
The sequential perfect-fit rule presented here (i.e., $\tau=0$) achieves
asymptotically the same rate. For example, Fig.~\ref{fig:scalingL} shows
that for $K=10$, $p=0.5$, and $L= 50\, 000$, the global coding gain
achieved by the perfect-fit rule is more than $4.5$, which is quite
close to the limiting value of around $5.0$
resulting from~\eqref{eq:limit}. 

\begin{figure}[htbp]
    \centering 
    \hspace{-0.3cm}\includegraphics{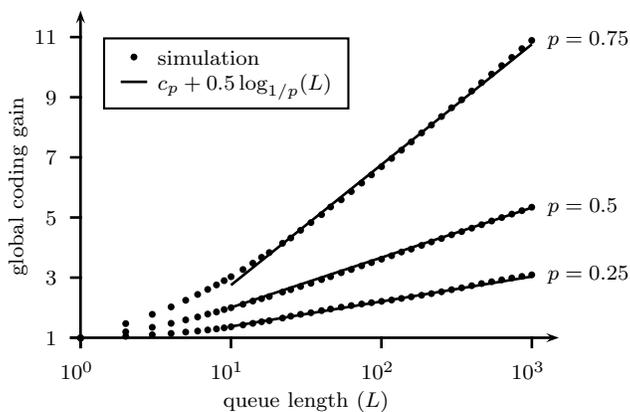}
    \caption{Scaling of the global coding gain as a function of the
        queue length $L$ in the system for first-fit merging ($\tau =K-2$)
         and different values of cache probability $p$.
        In the figure, the queue length $L$ is equal to $K$, and
        $c_{0.25} = 0.55$, $c_{0.5} = 0.35$, $c_{0.75} = -1.25$.  }
    \label{fig:scalingK}
\end{figure}

It is also instructive to consider the behavior of the global coding
gain as the length of the queue increases. This behavior is depicted in
Fig.~\ref{fig:scalingK} for the scenario in which the queue length $L$
is equal to the number of caches $K$. This scaling of $L$ with $K$
corresponds to each user tolerating only a delay of a single outstanding
symbol. Fig.~\ref{fig:scalingK} shows that even in this setting with
extremely tight deadline constraints, the global coding gain can be quite
substantial, being as large as a factor $11$ for $L=1000$ queue length under a
first-fit merging rule (i.e., $\tau=L-2$). Results from coloring of
large random graphs~\cite{grimmett75} suggest that we should expect the
global coding gain to scale approximately as $(0.5\pm
o(1))\log_{1/p}(L)$ in this scenario. Fig.~\ref{fig:scalingK} indicates
that this seems indeed to be the case.

\section{Experimental Results}
\label{sec:implementation}

\begin{figure}[htbp]
    %\centering 
    \hspace{5.2cm}\includegraphics{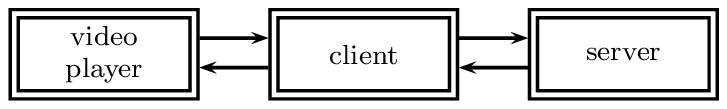}
    \caption{System architecture.} 
    \label{fig:schematic}
\end{figure}

To test the ideas developed in this paper in practice, we developed a
video-streaming prototype making use of the coded caching approach. The
prototype consists of three components: a video player, a client, and a
server, as depicted in Fig.~\ref{fig:schematic}. 

The server implements the operations of the origin server of the system.
The server reads requests for content symbols from the client and
responds by sending coded content symbols back to the client. The client
implements the operations of the cache. The client reads requests for
video content from the video player, converts these video requests into
requests for a sequence of content symbols, and requests from the server
those content symbols not found in its local cache. The client then
reads the coded content symbols sent by the server, decodes them with
the help of its local cache, reorders the decoded content symbols, and
forwards them in order to the video player. Finally, the video player
interacts with the human end user. The player transmits a video request
to the client, reads the corresponding response, and displays it to the
user. To simulate $K$ caches, we replicate the player and client
processes $K$ times.

Conceptually, the video player, client, and server processes are
placed at different points in the network (at the user, the edge cache,
and the origin server, respectively). However, for simplicity, two or
more of them may run on the same machine. For the video player, we use the
open-source VLC media player. The client and server are implemented in
Java; the client is about $500$ lines of code, the server about $1000$,
with an additional about $500$ lines of shared code.

The content database consists of three open-source videos (``Elephants
Dream'', ``Big Buck Bunny'', and ``Sintel'') from the Blender
Foundation, transcoded into Flash Video (FLV) format. The files have a
size of around $60$ MB each, corresponding to a bit rate of
approximately $700$ kb/s. Each video stream is divided into content
symbols of constant size $10$ kB. A uniquely identifying header is
attached to each symbol. This header consists of a string identifying
the video (``video1.flv'', ``video2.flv'', ``video3.flv'') together with
a $32$-bit sequence number. For coded content symbols, the header
consists of the number of symbols combined into this coded symbol
together with a list of corresponding symbol headers. Compared to the
symbol size of $10$ kB, the size of the headers is small.

The client caches are populated offline. We simulate this procedure by
reading the entire video database from disk, but keeping only a fraction
$p$ of read content symbols in main memory. Which symbols to keep is
decided using a pseudo-random number generator initialized with a
randomly chosen seed. The seed value is sent from the client to
the server at the beginning of the communication protocol. The server
recomputes the cache configuration of the client by using the same
pseudo-random number generator initialized with identical seed.

All communication between the three components of the video-streaming
prototype uses TCP. To keep the number of components needed to a
minimum, we simulate application-layer multicast from the server to the
clients by using several parallel unicast connections over which
identical content is sent.

\begin{figure}[htbp]
    %\centering 
    \hspace{4.5cm}\includegraphics{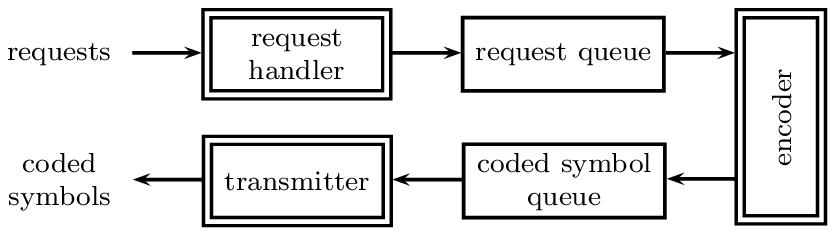}
    \caption{Schematic of server process. } 
    \label{fig:schematic_server}
\end{figure}

The most interesting aspect of the implementation (and the focus of
Sections~\ref{sec:problem} and~\ref{sec:main}) is the server, which
consists of three main components as shown in
Fig.~\ref{fig:schematic_server}. A number of request-handler threads,
one for each client, read symbol requests sent by the clients over the
network. New requests are enqueued in the request queue (implemented as
a Java LinkedBlockingQueue). A single encoder thread dequeues requests
from the request queue. It then traverses the coded symbol queue
(implemented as a Java DelayQueue) to find an existing request with
which the new request can be merged using a $\tau$-fit threshold merging
rule. If the new request cannot be merged with any existing request, it
is enqueued in the coded symbol queue ordered by time to deadline. The
transmitter thread dequeues from the coded symbol queue elements that
are close to their deadline and ``multicasts'' them over the network to
the clients.

\begin{figure}[htbp]
    \centering 
    \scalebox{0.35}{\includegraphics[clip=true,trim=0 2cm 3cm 0]{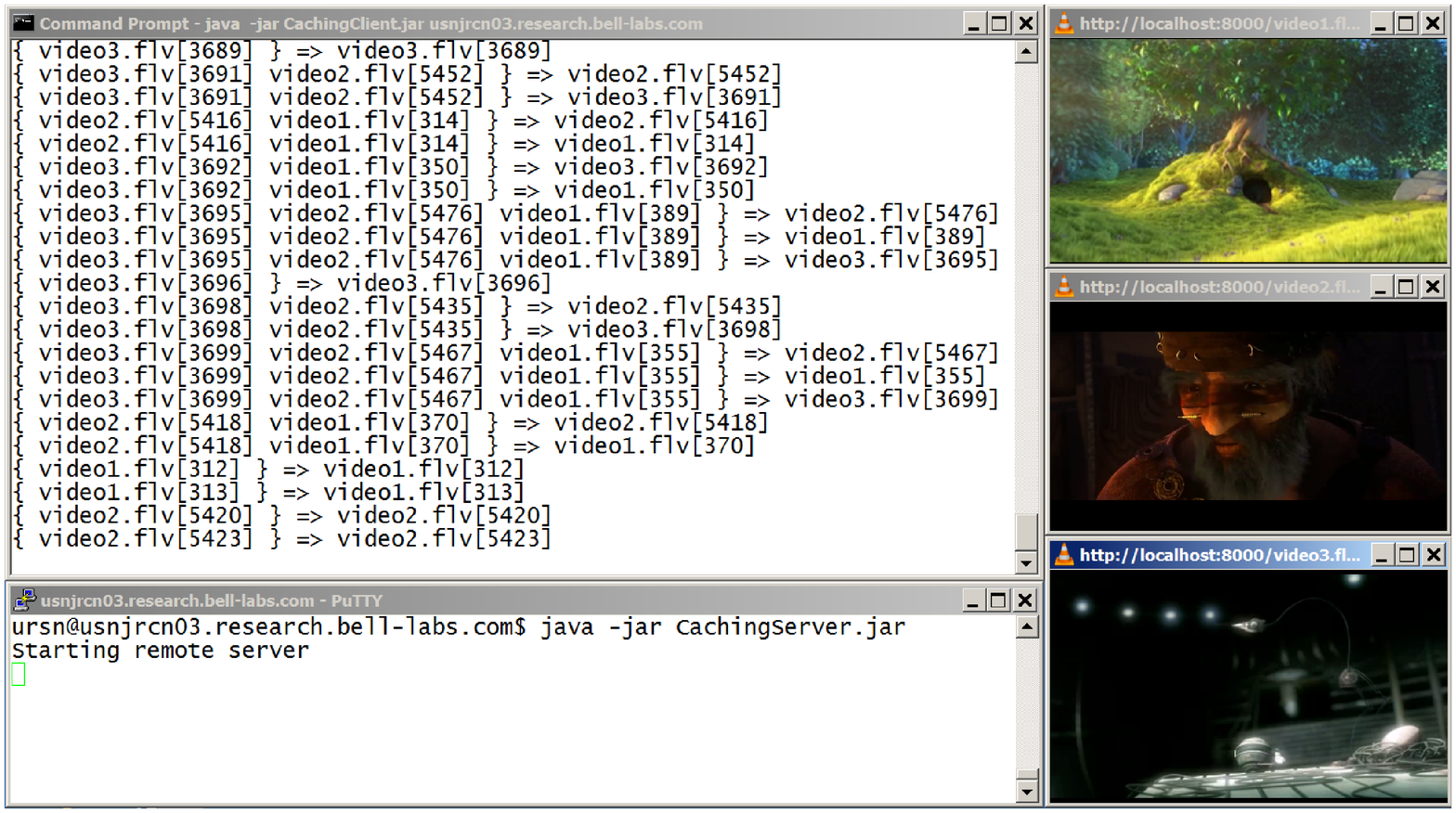}}
    \caption{Screenshot of video-streaming demo. The server process runs
        on a remote machine (usnjrcn03.research.bell-labs.com); the
    client and video player processes run locally.} 
    \label{fig:demo}
\end{figure}

Fig.~\ref{fig:demo} shows a screenshot of our video-streaming prototype,
in which three users simultaneously request a different video each
(right of screen). The bottom-left window shows the server process
handling the encoding of the video streams. The top-left window shows
the client processes handling the decoding and reconstruction of the
video streams. For example, in line $4$, the client indicates that it
received a coded symbol containing the XOR of symbol $5416$ of video
stream two and symbol $314$ of video stream one, from which it
decoded the symbol of stream two. Similarly, line $8$ shows the
decoding operation for a coded symbol consisting of the XOR of three
symbols from different video streams. 

\begin{figure}[htbp]
    \centering 
    \hspace{-0.1cm}\includegraphics[clip=true,trim=0 2.8cm 0 0]{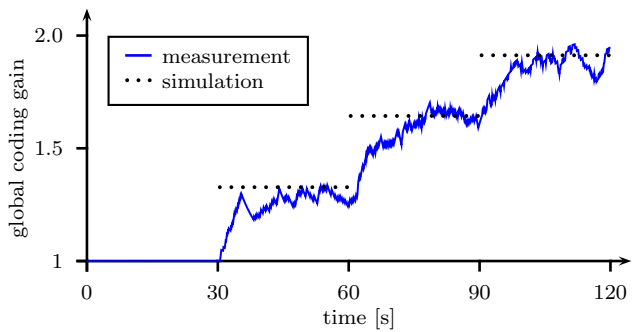}
    \caption{Global coding gain for video-streaming demo. In this
        trace, a first-fit threshold rule ($\tau=K-2$) is used, caches
        hold half of the content in the database ($p=0.5$), and the
        delay requirement is $50$ outstanding symbols at each user
        ($L\approx 50K$). The trace is exponentially smoothed with an effective 
    window size of roughly $40$ measurements.}
    \label{fig:coding_gain}
\end{figure}

The performance of our implementation as measured by the global coding
gain is depicted in Fig.~\ref{fig:coding_gain}. The figure shows a $120$
s trace for a number of concurrent users increasing by one every
$30$ s from one to four. As can be seen from the figure, the global
coding gain increases with the number of concurrent requests being
handled. For the last $30$ s of the trace, during which four videos are
served simultaneously, the coding gain is almost $2$, corresponding to a
server bandwidth reduction of close to $50\%$ compared to uncoded
transmission. The coding gain of the trace is close to the simulated
values (also shown in the figure). Because the coding gain reported from
the trace takes overhead into account, whereas the coding gain from the
simulations does not, this result confirms that the overhead is negligible.

\section{Discussion}
\label{sec:discussion}

The caching problem is related to the index-coding problem~\cite{birk06,
bar-yossef11} (or, equivalently~\cite{effros12}, to the network-coding
problem~\cite{ahlswede00}). In the index-coding problem, a server is
connected to users through a shared link. The server has access to a
database of files. Each user has local access to a fixed subset of these
files and is interested in one file not locally available. The objective
is to deliver the requested files to the users with the minimum number
of transmissions from the server. From this description, we see that for
\emph{fixed} cache content and for \emph{fixed} requests, the caching
problem induces an index-coding subproblem.

The caching problem considered in this paper and the index-coding
problem differ as follows. First, in the index-coding problem the
locally available files are fixed, whereas in the caching problem they
need to be designed. In other words, choosing the side information is
part of the problem statement in index coding, but part of the solution
in the caching problem. Second, in index-coding, the files requested by
the users are fixed, whereas in caching any of exponentially many
requests are possible. Therefore, after having deciding on the cache
content, we are faced with exponentially many index-coding subproblems
in the caching problem. Third, the notion of delay, which is a central
quantity in this work, is absent in index coding.

\section*{Acknowledgment}

The authors thank S.~Stolyar and M. Carroll for helpful discussions.

\end{document}